\documentclass[letter,twocolumn,10pt,twoside]{asme2e}

\confshortname{DSCC 2011}
\conffullname{2011 ASME Dynamic Systems and Control Conference}
\confdate{31}
\confmonth{October}
\confyear{2011}
\confcity{Arlington, VA}
\confcountry{USA}

\papernum{DSCC2011-5923}

\usepackage{amsmath, amssymb, amsxtra, mathrsfs}
\usepackage{graphics, epsfig, here, fancyhdr,graphicx}
\usepackage{bbm}
\usepackage[all]{xy}

\newcommand{\bN}{\mathbb{N}}

\newcommand{\bR}{\mathbb{R}}

\newcommand{\cJ}{\mathcal{J}}

\newcommand{\cO}{\mathcal{O}}

\newcommand{\ol}[1]{{\overline{#1}}}
\newcommand{\dS}{\displaystyle}

\newcommand{\bc}[1]{{\left\{#1\right\}}}
\newcommand{\bq}[1]{{\left[#1\right]}}
\newcommand{\bp}[1]{{\left(#1\right)}}

\newcommand{\bt}[1]{{\left\langle#1\right\rangle}}

\newcommand{\set}[2]{{\bc{#1\,\,:\,\,#2}}}

\newcommand{\vphi}{\varphi}
\newcommand{\Dx}{\dot{x}}

\renewcommand{\O}{\Omega}

\newcommand{\w}{\omega}
\newcommand{\cF}{\mathcal{F}}

\newcommand{\eP}{\varepsilon}

\newcommand{\wh}[1]{{\widehat{#1}}}

\newcommand{\half}{\frac{1}{2}}
\newcommand{\tH}{^{\mathrm{th}}}

\newcommand{\rd}{\mathrm{d}}

\title{Optimal Asymptotic Entrainment of Phase-Reduced Oscillators}

\author{Anatoly Zlotnik
    \affiliation{
    Department of Electrical and Systems Engineering\\
    Washington University\\
    Saint Louis, Missouri 63130\\
    Email: azlotnik@ese.wustl.edu
    }
}

\author{Jr-Shin Li
    \affiliation{
    Department of Electrical and Systems Engineering\\
    Washington University\\
    Saint Louis, Missouri 63130\\
    Email: jsli@ese.wustl.edu
    }
}

\begin{document}
\thispagestyle{headings}

\maketitle
\textfloatsep=20pt


\begin{abstract}
\textit{We derive optimal periodic controls for entrainment of a self-driven oscillator to a desired frequency.  The alternative objectives of minimizing power and maximizing frequency range of entrainment are considered.  A state space representation of the oscillator is reduced to a linearized phase model, and the optimal periodic control is computed from the phase response curve using formal averaging and the calculus of variations. Computational methods are used to calculate the periodic orbit and the phase response curve, and a numerical method for approximating the optimal controls is introduced.  Our method is applied to asymptotically control the period of spiking neural oscillators modeled using the Hodgkin-Huxley equations. This example illustrates the optimality of entrainment controls derived using phase models when applied to the original state space system.}
\end{abstract}

\section{INTRODUCTION}

The synchronization of oscillating systems is an important
and extensively studied phenomenon in science, and also finds
numerous engineering applications \cite{strogatz01}.  Examples
include the oscillation of neurons \cite{hoppensteadt97}, sleep
cycles and other pacemakers in biology
\cite{hanson78,mirollo90,ermentrout84}, semiconductor
lasers in physics \cite{fischer00}, and vibrating systems in
mechanical engineering \cite{blekhman88}.  The asymptotic
synchronization of an oscillator to a periodic control signal is
called entrainment,  and is studied  by examining the phase response
curve (PRC) \cite{izhikevich06,izhikevich07}, which
quantifies the shift in asymptotic phase due to an infinitesimal
perturbation in the state.  The classic phase coordinate
transformation \cite{malkin49} for studying nonlinear oscillators
was used together with formal averaging \cite{kornfeld82} to develop
a model of coupled chemical oscillations \cite{kuramoto84}.  Phase
models are widely used in physics, chemistry, and biology
\cite{pikovsky01} to study systems where the phase, but not the
state, can be observed, and where the PRC can be approximated
experimentally.  Interest in controlling synchronization in
electrochemical \cite{kiss02} and neural \cite{hoppensteadt99}
systems  has been increasing, and a method for approximating optimal
waveforms for entrainment of phase-reduced oscillators by weak
forcing has been proposed \cite{harada10}.

In this paper, we extend the theory of optimal entrainment of oscillators via weak, periodic controls \cite{harada10} to systems where the phase model has arbitrary PRC.  We also present an efficient numerical method that accurately computes optimal waveforms by finding the maximum of a polynomial whose coefficients depend on the PRC of the entrained oscillator.  This enables an examination of the important issue of how controls derived using the PRC perform when applied to entrain the associated oscillator in state space, which is the ultimate purpose of using phase models.  In the following section, we discuss the phase coordinate transformation for a nonlinear oscillator and the available numerical methods for computing the PRC, and describe how averaging theory is used to study the asymptotic behavior of oscillating systems.  In section 3, we use calculus of variations to derive theoretical entrainment controls that are optimal in the sense of minimum power or maximum entrainment range.  The former is optimal when the natural frequency of the entrained oscillator is known to be either above or below the desired value, and the latter is useful when the natural frequency is in a neighborhood of the desired value, but unknown.  We then present an efficient procedure for approximating these controls using Fourier series and Chebyshev polynomials.  Finally in section 4, our approach is employed to entrain the Hodgkin-Huxley neuron model.  The results suggest that optimal controls derived using a phase model are optimal for entrainment of the associated state space system.

\section{PHASE MODELS} \label{secpv}

Consider a smooth ordinary differential equation system
\begin{equation} \label{sys1}
\Dx=f(x,u), \quad x(0)=x_0,
\end{equation}
where $x(t)\in\bR^n$ is the state and $u(t)\in\bR$ is a control.  Furthermore, we require that (\ref{sys1}) has an attractive, non-constant limit cycle $\gamma(t)=\gamma(t+T)$, satisfying $\dot{\gamma}=f(\gamma,0)$, on the periodic orbit $\Gamma=\set{y\in\bR^n}{y=\gamma(t) \text{ for } 0\leq t< T}\subset\bR^n$.    In order to study the behavior of this system, we reduce it to a scalar equation
\begin{equation} \label{sys2}
\dot{\psi}=\w+Z(\psi)u,
\end{equation}
which is called a phase model, where $Z$ is the PRC and $\psi(t)$ is the phase associated to the isochron on which $x(t)$ is located. The isochron is the manifold in $\bR^n$ on which all points have asymptotic phase $\psi(t)$ \cite{brown04}.  The conditions for validity and accuracy of this model have been determined \cite{efimov10}, and the reduction is accomplished through the well-studied process of phase coordinate transformation \cite{efimov09}, which is based on Floquet theory \cite{perko90,kelley04}.  The model is assumed valid for inputs $u(t)$ such that the solution $x(t,x_0,u)$ to (\ref{sys1}) remains within a neighborhood of $\Gamma$.
To compute the PRC, the period $T=2\pi/\w$ and the limit cycle $\gamma(t)$ must be computed to a high degree of accuracy.  This is done using a method for determining the steady-state response of nonlinear oscillators \cite{aprille72} based on perturbation theory \cite{khalil02} and gradient optimization \cite{peressini00}.
The PRC can then be computed by integrating the adjoint of the linearization of (\ref{sys1}) \cite{ermentrout96}, or by using a more efficient and numerically stable spectral method developed more recently \cite{govaerts06}.  A software package called XPPAUT \cite{ermentrout02} is commonly used by researchers to compute the PRC.  We use a modified spectral method in our implementation that is very accurate for stiff systems.

Our goal is to entrain the system (\ref{sys2}) to a new frequency $\O$ using a periodic control $u(t)=k(\O t)$ where $k$ is $2\pi$-periodic.   We have adopted the weak forcing assumption, i.e. $k=\eP k_1$ where $k_1$ has unit power, so the original system (\ref{sys1}) is guaranteed to traverse a neighborhood of $\Gamma$ given this control.  Now define a slow phase variable by $\phi(t)=\psi(t)-\O t$, and call the difference $\Delta\w=\w-\O$ between the natural and forcing frequencies the frequency detuning.  The dynamic equation for the slow phase is
\begin{equation} \label{sys3}
\dot{\phi}=\dot{\psi}-\O=\Delta\w+Z(\O t+\phi)k(\O t),
\end{equation}
where $\dot{\phi}$ is called the phase drift.  In order to study the asymptotic behavior of (\ref{sys3}) it is necessary to eliminate the dependence on time, which can be accomplished by using formal averaging \cite{kuramoto84}.  Given a periodic forcing with frequency $\O=2\pi/T$, we denote the forcing phase $\theta=\O t$.   If $\cF$ is the set of $2\pi$-periodic functions on $\bR$, we can define an averaging operator $\bt{\cdot}:\cF\to\bR$ by
\begin{equation} \label{ave}
\bt{x}=\frac{1}{2\pi}\int_0^{2\pi}x(\theta)\rd \theta.
\end{equation}
The weak ergodic theorem for measure-preserving dynamical systems on the torus \cite{kornfeld82} implies that for any $\phi$,
\begin{align} \label{Lambdef}
\Lambda(\phi) & = \bt{Z(\theta+\phi)k(\theta)} \notag\\ & =
\frac{1}{2\pi}\int_0^{2\pi} Z(\theta+\phi)k(\theta)\rd \theta \\ & =\lim_{T\to\infty}\frac{1}{T}\int_0^TZ(\O t+\phi)k(\O t)\rd t \notag
\end{align}
exists as a smooth, $2\pi$-periodic function in $\cF$.  By the formal averaging theorem \cite{hoppensteadt97}, the system
\begin{equation} \label{sys4}
\dot{\vphi}=\Delta \w + \Lambda(\vphi)+\cO(\eP^2)
\end{equation}
approximates (\ref{sys3}) in the sense that there exists a change of variables $\vphi=\phi+\eP h(\vphi,\phi)$ that maps solutions of (\ref{sys3}) to those of (\ref{sys4}).  Therefore the weak forcing assumption $k=\eP k_1$ with $\eP<<1$ allows us to approximate the phase drift equation by
\begin{equation} \label{sys5}
\dot{\vphi}=\Delta \w + \Lambda(\vphi).
\end{equation}
The averaged equation (\ref{sys5}) is independent of time, and can be used to study the asymptotic behavior of the periodically forced system (\ref{sys2}) where $u=k(\O t)$.

\section{ENTRAINMENT OF PHASE MODELS}

We call the system (\ref{sys2}) entrained by a control $u=k(\O t)$ when the phase drift equation (\ref{sys5}) satisfies $\dot{\vphi}=0$. This occurs when there exists a phase $\vphi_*$ satisfying $\Delta \w + \Lambda(\vphi_*) = 0$, in which case the system is called entrainable. Defining the phases $\vphi_-=\arg\min_\vphi \Lambda(\vphi)$ and $\vphi_+=\arg\max_\vphi\Lambda(\vphi)$, we can formulate entrainment as an optimal control problem.  When the objective is to minimize the control power $\bt{k^2}$, entrainability requires that
\begin{equation} \label{const1}
\begin{array}{rcl}
\Delta \w+\Lambda(\vphi_+)=0  &\quad \text{if}\quad & \O>\w,\\\Delta \w+\Lambda(\vphi_-)=0 &\quad\text{if}\quad& \O<\w.
\end{array}
\end{equation}
We formulate the problem for $\O>\w$, and the case where $\O<\w$ is symmetric.
The constraint (\ref{const1}) can be added by adjoining it to the objective function using a multiplier $\lambda$, resulting in
\begin{eqnarray} \label{op1}
\min \cJ[k] & = & \bt{k^2} - \lambda(\Delta\w + \Lambda(\vphi_+)) \\ & = &
\bt{k^2} - \lambda\bp{\Delta\w + \frac{1}{2\pi} \int_0^{2\pi} Z(\theta+\vphi_+)k(\theta)\rd \theta} \notag\\
& = & \frac{1}{2\pi} \int_0^{2\pi} \bq{k(\theta)(k(\theta)-\lambda Z(\theta+\vphi_+)) - \lambda\Delta \w}\rd \theta \notag
\end{eqnarray}
The Euler-Lagrange equation provides necessary conditions for the optimal solution, which is given by
\begin{equation*}
k_*(\theta)=\frac{\lambda}{2} Z(\theta+\vphi_+).
\end{equation*}
The constraint (\ref{const1}) can be used to solve for $\lambda$, because
\begin{equation}
0 = \Delta \w+\Lambda_*(\vphi_+) = \Delta \w + \frac{1}{2\pi}
\int_0^{2\pi} \frac{\lambda}{2} Z(\theta+\vphi_+)^2 \rd \theta
\end{equation}
implies that $\lambda=-2\Delta \w/\bt{Z^2}$.  Consequently the minimum power control is
\begin{equation} \label{sol1}
k_*(\theta)= -\frac{\Delta \w}{\bt{Z^2}} Z(\theta),
\end{equation}
with power $P=(\Delta \w)^2/\bt{Z^2}$.  We omit the phase ambiguity $\vphi_+$ in the solution $k_*$ because entrainment is asymptotic.

Now consider the dual problem where for fixed power $P$, a periodic
waveform $k(\O t)$ is derived to maximize the locking range $R[k]$
of natural frequencies $\w$ for which the family of oscillators
$\set{\dot{\psi}=\w+Z(\psi)u}{\w\in(\w_{\min},\w_{\max})}$ can be
entrained to a forcing frequency $\O$ \cite{harada10}.   The locking
range is given by $R[k]=\w_{\max}-\w_{\min} = \Delta\w_{\min} -
\Delta\w_{\max} = \Lambda(\vphi_+)-\Lambda(\vphi_-)$, so that
adjoining the constraint on the power to the objective function
using a multiplier $\lambda$ gives rise to the optimal control
problem
\begin{eqnarray} \label{op2}
\cJ[k;P] & = & R[k]-\lambda(\bt{k^2}-P) \\ & = & \Lambda(\vphi_+)-\Lambda(\vphi_-) - \lambda(\bt{k^2}-P) \notag \\
& = & \bt{Z(\theta+\vphi_+)k(\theta)}-\bt{Z(\theta+\vphi_-)k(\theta)}-\lambda(\bt{k^2}-P) \notag\\
& = & \dS \frac{1}{2\pi}\int_0^{2\pi} \bp{k(\theta)[Z(\theta+\vphi_+)-Z(\theta+\vphi_-)-\lambda k(\theta)] + \lambda P} \rd \theta \notag
\end{eqnarray}
Solving the Euler-Lagrange equation yields
\begin{equation*}
k_*(\theta)=\dS\frac{1}{2\lambda}[ Z(\theta+\vphi_+)-Z(\theta+\vphi_-)].
\end{equation*}
The optimal solution $k_*$ satisfies the constraint
$\bt{k_*^2}-P=0$, so
$$
\frac{1}{2\pi}\int_0^{2\pi} \bp{\frac{1}{2\lambda}}^2 [ Z(\theta+\vphi_+)-Z(\theta+\vphi_-)]^2\rd \theta - P =0,
$$
and hence $\lambda=\frac{1}{2}\sqrt{Q/P}$ where $Q=\bt{[
Z(\theta+\vphi_+)-Z(\theta+\vphi_-)]^2}$.  Substituting this into
(\ref{Lambdef}) gives
\begin{align} \label{Lambopt}
\Lambda_*(\vphi) & =  \bt{Z(\vphi+\theta)k_*(\theta)} \notag \\ & =  \dS\frac{1}{2\lambda}\bt{Z(\vphi+\theta)[Z(\theta+\vphi_+)-Z(\theta+\vphi_-)]} \notag\\
& =  \sqrt{P/Q}\bt{Z(\vphi+\theta)[Z(\theta+\vphi_+)-Z(\theta+\vphi_-)]}.
\end{align}
Because $Z(\theta)$ is $2\pi$-periodic, we represent it as a Fourier series,
\begin{equation} \label{prcf}
\dS Z(\theta)= \dS \half a_0 + \sum_{n=1}^\infty a_n\cos(n\theta) + \sum_{n=1}^\infty b_n\sin(n\theta),
\end{equation}
and we find that for $\vphi_1,\vphi_2\in[0,2\pi)$,
\begin{equation} \label{prcinp}
\bt{Z(\vphi_1+\theta)Z(\vphi_2+\theta)}=\dS \frac{1}{4}a_0^2 + \half \sum_{n=1}^\infty (a_n^2+b_n^2) \cos(n(\vphi_1-\vphi_2)).
\end{equation}
Substituting this result into (\ref{Lambopt}), we obtain
\begin{equation} \label{Lamopt}
\Lambda_*(\vphi) = \dS \sqrt{\frac{P}{4Q}}\sum_{n=1}^\infty (a_n^2+b_n^2)[\cos(n(\vphi-\vphi_+))-\cos(n(\vphi-\vphi_-))].
\end{equation}
Let us denote the phase difference $\Delta \vphi = \vphi_+-\vphi_-$.  Then
\begin{align} \label{Qfun}
Q & = \bt{[ Z(\theta+\vphi_+)-Z(\theta+\vphi_-)]^2} \notag \\
 & =  \bt{Z(\theta+\vphi_+)^2} - 2\bt{Z(\theta+\vphi_+)Z(\theta+\vphi_-)} +  \bt{Z(\theta+\vphi_-)^2} \notag\\
 & =  \dS \sum_{n=1}^\infty (a_n^2+b_n^2)[1-\cos(n\Delta\vphi)].
\end{align}
By substituting $\vphi_-$ and $\vphi_+$ into (\ref{Lamopt}), we obtain the optimal locking range $R[k_*]$ as a function of $\Delta \vphi$ and the Fourier coefficients of $Z$, namely
\begin{align} \label{rang}
R[k_*] & =\Lambda_*(\vphi_+)-\Lambda_*(\vphi_-) \\ & =  \sqrt{P/Q}\sum_{n=1}^\infty (a_n^2+b_n^2)[1-\cos(n\Delta\vphi)] \,\,\, = \,\,\, \sqrt{P}\sqrt{Q}. \notag
\end{align}
Consequently, to find the optimal control $k_*$ and the maximum locking range $R[k_*]$, it suffices to maximize $Q$ in terms of $\Delta \vphi$. The value of $\Delta \vphi$ that maximizes $Q$ in (\ref{Qfun}) also satisfies the first order condition $Q'(\Delta\vphi) = \sum_{n=1}^\infty n(a_n^2+b_n^2)\sin(n\Delta\vphi)=0$, hence there exists a ``generic'' solution $\Delta \vphi=\pi$, which may not be optimal.  Observe that if we set $y=\cos(\Delta\vphi)$, then
\begin{equation} \label{qcrit}
Q(\Delta\vphi)=q(y) =  \sum_{n=1}^\infty(a_n^2+b_n^2)[1-T_n(y)],
\end{equation}
where $T_n$ is the $n\tH$ Chebyshev polynomial of the first kind.  Therefore a straightforward criterion for the existence of superior solutions is to check whether $q$ attains its supremum on $(-1,1)$. In that case we choose $\Delta \vphi =\pm\arccos(y_*)$, and otherwise we choose $\Delta \vphi = \pi$.  The optimal waveform is given by
\begin{equation} \label{sol2}
k_*(\theta)=\sqrt{P/Q}[ Z(\theta+\Delta\vphi)-Z(\theta)].
\end{equation}
We omit the phase ambiguity $\vphi_-$ in (\ref{sol2}) because entrainment is asymptotic. The two possible values for $\Delta \phi$ result in two optimal solutions when the criterion for (\ref{qcrit}) holds.

\section{ENTRAINMENT OF NEURONS}

The notion of modeling the dynamics of neurons in the human brain as oscillators has gained wide acceptance among researchers in neuroscience and mathematical biology \cite{brown04,moehlis06}.  Because the ability to control the synchronization of neural dynamics has important research and clinical implications \cite{good09,schiff94}, it is important to explore the pertinence of the entrainment paradigm to neural systems.  We consider the entrainment of a neuron by an external stimulus, and use as an example the model of Hodgkin and Huxley \cite{hodgkin52}.  Starting with the commonly used parameterization \cite{brown04}, we reduce the system to the phase model and compute optimal entrainment controls.   The objective is either to entrain the model to a given frequency with minimum power (\ref{op1}), or to maximize the range of frequencies (and hence the number of neurons) that can be entrained by a control of fixed power (\ref{op2}).  For a given waveform $k(\O t)$ where $\O$ is in a neighborhood of the natural frequency $\w$, we can numerically approximate the power actually required for entrainment.  This allows us to compute the  approximately triangular region of entrainability called the Arnold tongue, which is the plot of the minimum amplitude $\sqrt{P}$ required for entrainment versus forcing frequency $\O$, and which is commonly used to visualize the asymptotic properties of an oscillating system \cite{pikovsky01,coombes99}. This will be used to illustrate the performance of the controls that we have derived.

\begin{figure}
\centerline {\includegraphics[height=4 cm, width=1.1\linewidth]{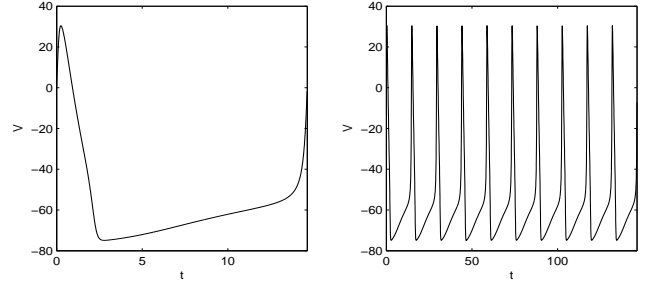}
} \vskip-.2cm\caption{Hodgekin-Huxley limit cycle (left) and ``spiking'' (right)} \label{fig1} \end{figure}

\begin{figure}
\centerline { \includegraphics[height=4 cm, width=1.1\linewidth]{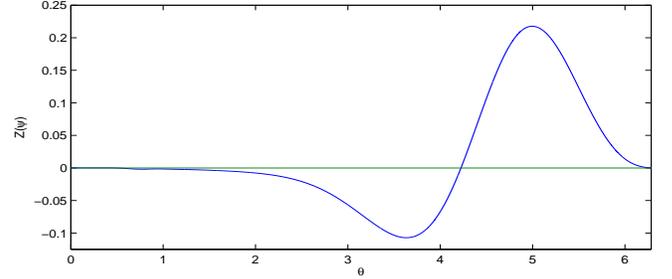}
} \vskip-.2cm\caption{Hodgekin-Huxley phase response curve (PRC)} \label{fig2} \end{figure}

The Hodgkin-Huxley model describes the propagation of action potentials in neurons, specifically the squid giant axon, and is used as a canonical example of neural oscillator dynamics. The equations are
\begin{equation} \label{hheq}
\hskip-2pt\begin{array}{c}\begin{array}{rcl}
c\dot{V}&=&I_b+I(t)-\ol{g}_{Na}h(V-V_{Na})m^3-\ol{g}_K(V-V_k)n^4-\ol{g}_L(V-V_L)\\
\dot{m}&=&a_m(V)(1-m)-b_m(V)m,\\
\dot{h}&=&a_h(V)(1-h)-b_h(V)h,\\
\dot{n}&=&a_n(V)(1-n)-b_n(V)n,
\end{array}\\\\\begin{array}{rcl}
a_m(V)&=&0.1(V+40)/(1-\exp(-(V+40)/10)),\\ b_m(V)&=&4\exp(-(V+65)/18),\\
a_h(V)&=&0.07\exp(-(V+65)/20),\\ b_h(V)&=&1/(1+\exp(-(V+35)/10)),\\
a_n(V)&=&0.01(V+55)/(1-\exp(-(V+55)/10)),\\ b_n(V)&=&0.125\exp(-(V+65)/80).\\
\end{array}\end{array}
\end{equation}
The variable $V$ is the voltage across the axon membrane, and $m$, $h$, and $n$ are the ion gating variables.  $I_b$ is a baseline current that induces the oscillation, and $I(t)$ is the control input.  The units of $V$ are millivolts and the units of time are milliseconds. We analyze this system of differential equations as an oscillator $\Dx=f(x,u)$, with a periodic limit cycle $\gamma(t)=\gamma(t+T)$ present when $u\equiv 0$.  Using the standard parameters $V_{Na} =50 \text{ mV}$, $V_K=-77 \text{ mV}$, $V_L=-54.4 \text{ mV}$, $\ol{g}_{Na}=120 \text{ mS/cm}^2$, $\ol{g}_K = 36 \text{ mS/cm}^2$, $\ol{g}_L=0.3 \text{ mS/cm}^2$, $I_b=10 \,\,\mu\text{A/cm}^2$, and $c=1 \,\,\mu\text{F/cm}^2$, we compute the limit cycle, which is shown for the voltage $V$ in Figure \ref{fig1}.  The period is computed as $T=14.63842\pm10^{-5}$ ms. The ``spiking'' behavior of the oscillator indicates that this system is stiff, and hence ill-conditioned for numerical integration.  We use a second order Adams-Bashforth solver to integrate these equations with a relative error tolerance of $10^{-6}$.    The PRC is computed along the limit cycle with an initial condition $x_0=(V_0,m_0,n_0,h_0)=(0,0.51916,0.2999,0.4812)$ corresponding to $\psi(0)$, and the result is shown in Figure \ref{fig2}.  An absolute error lower than $10^{-4}$ is maintained by using a grid with step size $0.002$.  The first and second zero crossings occur at $\psi=0.4617$ and $\psi=4.2242$, respectively.  Note that $u$ is least effective at the start of the cycle, when the neuron is spiking.

\begin{figure}
\centerline { \includegraphics[height=4 cm, width=1.1\linewidth]{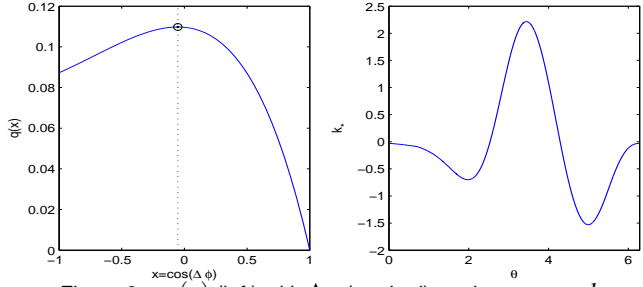}
} \vskip-.2cm\caption{$q(x)$ (left) with $\Delta \vphi$ (marked), and max range $k_*$} \label{fig3} \end{figure}

We expand the PRC in a Fourier series as in (\ref{prcf}) by using
the discrete Fourier transform of $\set{Z(j)}{j=2\pi j/N,\,\,
j=1,\ldots 5000}$ to approximate the coefficients.  This gives us
$\wh{Z}(n)=\sum_{j=1}^NZ(j)\w_N^{(j-1)(k-1)}$ where $\w_N=e^{-2\pi
i/N}$, and the estimates are $a_n=\Re(\wh{Z}(n))\cdot 2/N$ and
$b_n=-\Im(\wh{Z}(n))\cdot 2/N$. Because of the phase ambiguity, the
choice of $x_0\in\Gamma$ that is used to compute $\gamma(t)$
influences the values of $a_n$ and $b_n$, but not the value of
$|a_n+ib_n|$.  We take 20 Fourier modes for our approximation. The
total power of the Hodgkin-Huxley PRC as a periodic waveform is
$0.0387$, and the modes $k=1,2,\ldots,5$ have power 0.01706,
0.01649, 0.00473, 0.00048, and 0.00001, respectively. The modes $2$
and $3$ have significant power, hence it is insufficient to use a
single mode to approximate the PRC.  The minimum power waveform
(\ref{sol1}) is a re-scaled PRC.  To compute the maximum range
waveform (\ref{sol2}), we find that a value of $y_*=-0.05287$
maximizes the polynomial $q$ in (\ref{qcrit}) on $(-1,1)$, hence the
``generic'' solution $\Delta \vphi=\pi$ is not optimal, so we use
$\Delta \vphi=\arccos(x_*)=1.623690$ and get $Q\approx 0.10976\pm1$.
The polynomial $q$, its maximum, and the maximum range control
waveform (\ref{sol2}) with unity power are shown in Figure
\ref{fig3}.

\begin{figure}
\centerline { \includegraphics[height=5 cm,
width=1.1\linewidth]{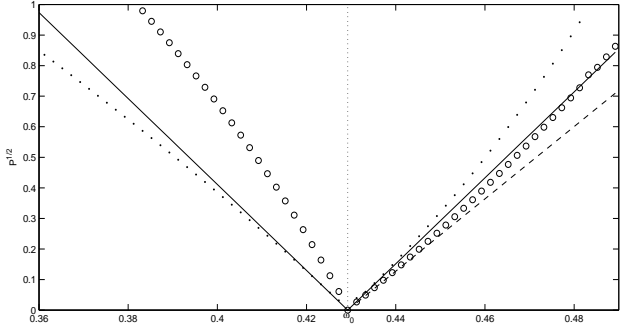} } \vskip-.2cm\caption{Arnold tongues
for Hodgkin-Huxley phase model (\ref{sys2}): Minimum power theory
(dashed line) and computation (o); Maximum range theory (solid line)
and computation ($\cdot$). The minimum power control functions as
intended only to increase frequency, while the maximum range control
has a useful symmetry property.} \label{figd}\end{figure}

\begin{figure}
\centerline { \includegraphics[width=1.2\linewidth]{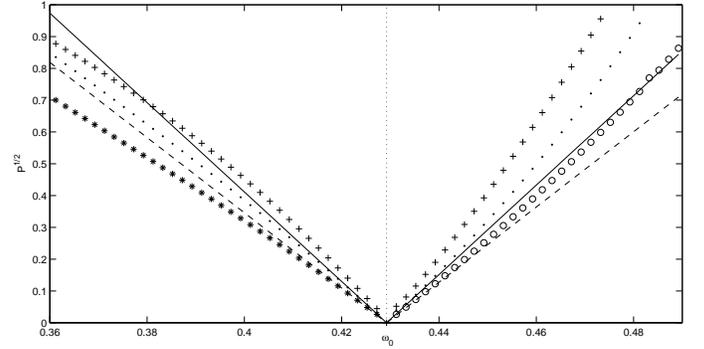} }
\vskip-.2cm\caption{Arnold tongue for Hodgkin-Huxley phase model
(\ref{sys2}): Minimum power theory (dashed line) and computation for
increase (o) and decrease ($*$) of frequency; Maximum range theory
(solid line) and computation ($\cdot$); sine wave computation ($+$).
The minimum power waveform for increasing (o) (decreasing ($*$))
$\w$ matches the theory (dashed line) closely near $\w_0$ for
$\w>\w_0$ ($\w<\w_0$). Similarly, the maximum range waveform
($\cdot$) matches the theory (solid line) closely near $\w_0$, and
can be effectively applied to increase or decrease the frequency.
The sine wave ($+$) has the worst performance.}
\label{fig4}\end{figure}

\begin{figure}
\centerline { \includegraphics[height=6 cm,
width=1.2\linewidth]{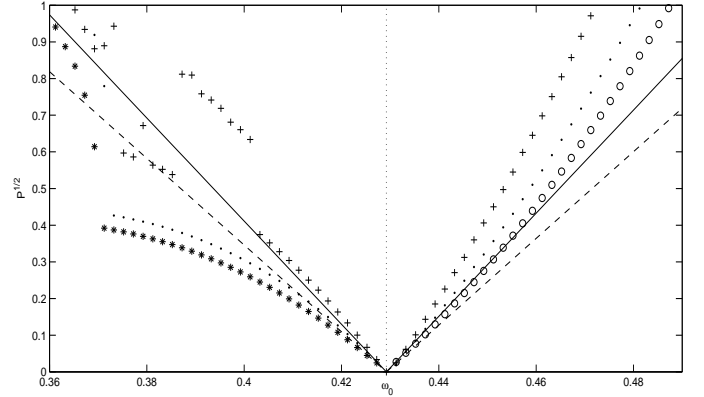} } \vskip-.2cm\caption{Arnold tongue
for Hodgkin-Huxley state-space model (\ref{hheq}): Computed minimum power
control for increase (o) and decrease ($*$) of frequency, and theory (dashed line); Computed maximum
range ($\cdot$), and theory (solid line); sine wave ($+$).} \label{fig5}\end{figure}

\newpage
 To evaluate the entrainability of a phase-reduced system by a given
waveform, we compute the Arnold tongue by determining the power
required for entrainment at a given frequency $\O$.  The key idea is
that if entrainment does indeed occur, then the response of the
oscillator is periodic with a period equal to $T=2\pi/\O$.  If the
solution to (\ref{sys2}) with $u=k(\O t)$ is sampled at this
interval and the sequence $\{\psi(jT)\}_{j\in\bN}$ converges, it
follows that the control $u$ entrains the phase model.  We determine
the power $P_*(\O)$ required for the sequence to converge by
performing a bisection search, using 150 points of the sequence as a
test.  A plot of $\sqrt{P_*(\O)}$ vs. $\O$ generates the resulting
Arnold tongue. The distinction between the solutions (\ref{sol1})
and (\ref{sol2}) obtained by using the alternative objectives is
illustrated in Figure \ref{figd}.  The results for (\ref{sol1}) on
the irrelevant range are omitted in other figures. The Arnold
tongues for the phase reduced system are presented in Figure
\ref{fig4}. Note that the actual Arnold tongues are not linear, and
the required power to decrease (increase) the frequency is lower
(higher) than predicted by the theory.  An issue of fundamental
importance is how well the entrainment control works when it is
applied to the original Hodgkin-Huxley system.  Figure \ref{fig5}
shows $\sqrt{P_*(\O)}$ vs. $\O$ when the same control waveforms are
applied to the original system (\ref{hheq}). The power required to
entrain the state space model to a frequency $\w$ is similar
to the theoretical prediction near the natural frequency.  By comparing Figures \ref{fig4} and \ref{fig5}, one sees that the relative entrainability of the phase
and state models by the tested waveforms is nearly identical for values of $\O$ near the natural frequency $\w_0$. This is strong evidence that optimal entrainment waveforms for a
phase-reduced oscillator (\ref{sys2}) are optimal in the same sense
for the state-space system (\ref{sys1}) from which the reduced model
is derived.

\section*{Conclusions}

We have presented a method for optimal entrainment of oscillators given the alternative objectives of minimum control power and maximum range of entrainability.  The method that we derived is based on the phase response curve of the oscillator and formal averaging theory.  We examine the entrainment of phase-reduced Hodgkin-Huxley neurons as an example problem, and compute Arnold tongues to evaluate the effectiveness of our controls.  Their performance closely matches the theoretical bounds when the weak forcing requirement is fulfilled.  The optimal waveforms produce a similar result when applied to the original model, which suggests that optimal entrainment controls for a phase model are optimal for the original system, provided the oscillator remains within a neighborhood of its limit cycle.    This work provides a basis for evaluating the effectiveness of phase reduction techniques for the control of oscillating systems.  The approach described is of direct interest to researchers in chemistry and neuroscience, and may also be applied to vibration control in engineered systems.

\bibliographystyle{asmems4}
\bibliography{prc_bib}

\end{document}